\documentclass[10pt,times,mathptm,aps,article,twocolumn,showpacs,showkeys]{revtex4}

\usepackage[]{graphicx}
\usepackage{amssymb}
\usepackage{bm}

\begin{document}

\setcounter{page}{1}
{\it Published in} Optical Materials, {\bf 27} (2005) 1421-1425

\copyright  2004 Elsevier B.V. All rights reserved.

{\it Available online at}  {\bf www.sciencedirect.com}

doi:10.1016/j.optmat.2004.07.016

\title[]{\vspace{15mm} Nanostructuring Lithium Niobate substrates by focused ion beam milling.\vspace{5mm}}
\author{F. \surname{Lacour}}
\author{M. \surname{Courjal}}
\author{M.-P. \surname{Bernal}}
\email{maria-pilar.bernal@univ-fcomte.fr}
\thanks{Website:\url{http://www.femto-st.fr/fr/Departements-de-recherche/OPTIQUE/}}
\author{A. \surname{Sabac}}
\author{C. \surname{Bainier}}
\author{M. \surname{Spajer}}

\affiliation{\vspace{2mm} Institut FEMTO-ST, D\' epartement d'Optique P. M. Duffieux,
Universit\' e de Franche-Comt\' e, UMR 6174 CNRS, 25030 Besan\c con cedex, France
\vspace{2mm}}

\date[]{Received 18 May 2004; accepted 14 July 2004}

\begin{abstract}
\vspace{5mm} We report on two novel ways for patterning Lithium Niobate (LN) at submicronic scale by means of focused ion beam (FIB) bombardment. The first method consists of direct FIB milling on $LiNbO_3$ and the second one is a combination of FIB milling on a deposited metallic layer and subsequent RIE (Reactive Ion Etching) etching. FIB images show in both cases homogeneous structures with well reproduced periodicity. These methods open the way to the fabrication of photonic crystals on $LiNbO_3$ substrates.

\end{abstract}



\maketitle

\section{Introduction}
The recent development of integrated photonic crystals within planar waveguides can help implementing compact devices with fully integrable functions \cite{kamp04,zimmermann04,wu03}. In these devices the light is confined into the crystal by a classical waveguide construction.  Lithium Niobate (LN) is a suitable material for 2D photonic crystals because of its high refractive index. Moreover, its high electro-optic coefficient and its low optical losses make this material very adequate for optical communication systems. Therefore, the perspective of fabricating miniature electro-optical and all-optical LN devices is attracting the research on $LiNbO_3$ nanostructuring \cite{restoin03,foglietti03}. However, the obtention of good nanometric optical structures in $LiNbO_3$ continues to be a difficult task due to its well-known resistivity towards standard machining techniques like wet etching \cite{emis89}.

In this paper, we report on two alternative methods based on focused ion beam (FIB) bombardment to produce photonic band gap structures on $LiNbO_3$ substrates with a spatial resolution of $70nm$.  The high resolution and the ability to drill holes directly from the sample surface make FIB milling one of the best candidates for designing good optical quality patterns at submicrometer scale \cite{bostan03}. The only constraint is that the sample surface must be metallised and grounded to avoid charge accumulation. Firstly, we describe the method for directly etching the LN substrate by FIB milling through the metal. This method has been already employed to etch sub-micrometric one-dimensional structures in  $LiNbO_3$ \cite{yin99}. The second related method is based on RIE etching after FIB milling of the metal layer which behaves as a mask. The advantage of this alternative solution is a lower exposure time. Another expected advantage whould be a good replication of the mask shape in the whole hole depth. In both cases, the fabricated submicronic patterns are characterized by FIB imaging.

Before describing the two nanostructuring methods, we present the calculated conditions to obtain a photonic band gap for $LiNbO_3$ substrates.

\section{Numerical study}

\begin{figure}[!htb]
        \includegraphics[width=0.45\textwidth]{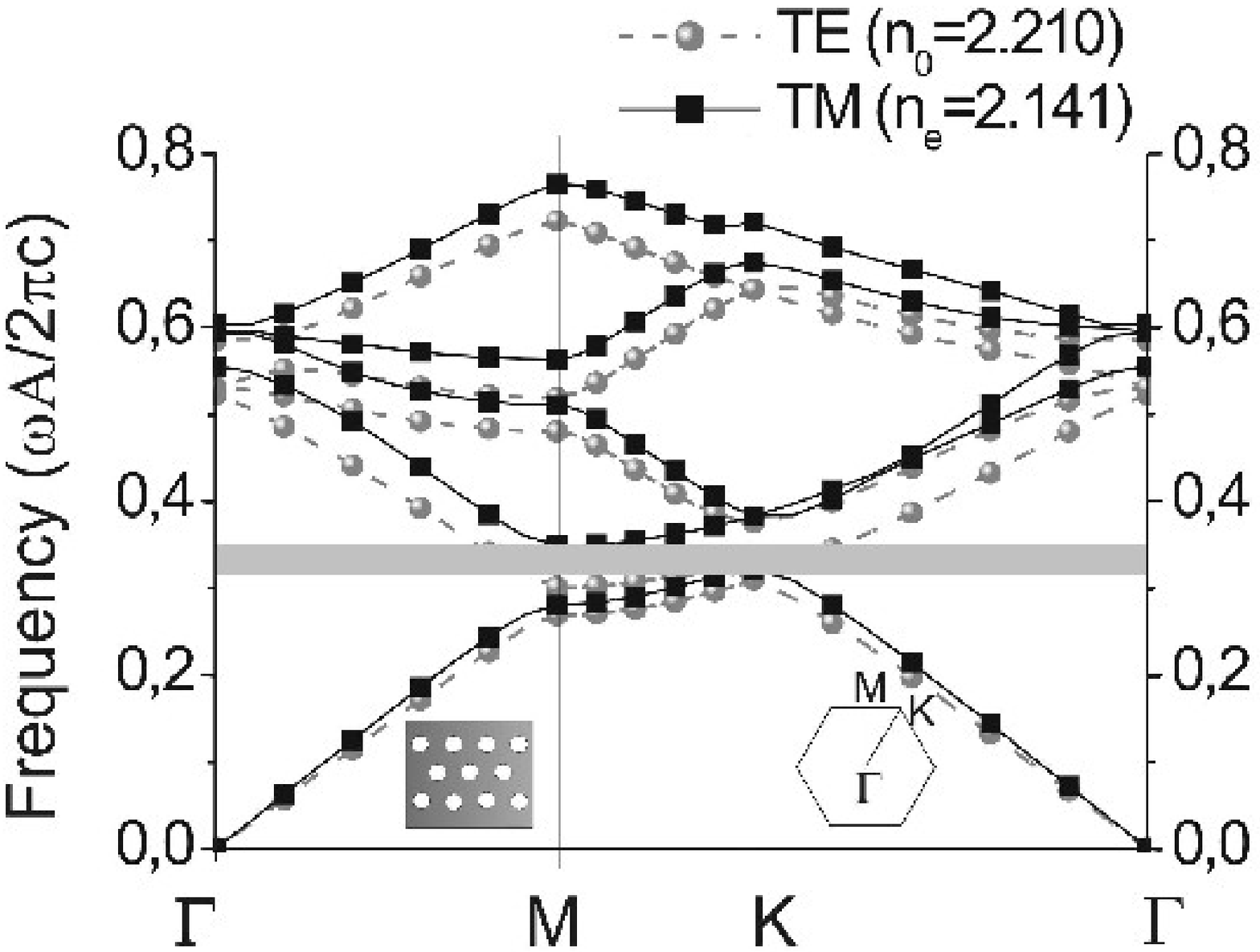}
        \caption{TE and TM band structure for 2D triangular array.}
        \label{fig.1}
\end{figure}

Numerical calculations are carried out using the commercial software {\it BandSolve}. For a wavelength $\lambda = 1.55 \mu m$, the ordinary and extraordinary indexes of the Z-cut $LiNbO_3$ substrate are assumed to be $n_o=2.2151$ and $n_e=2.1410$, respectively. According to our calculations, a 2D triangular lattice of holes can yield to a total TM photonic bandgap if the diameter $D$ of the holes is larger than $0.4 p$, where $p$ is the triangular lattice period. The ratio $D/p$ should be as large as possible in order to benefit from a large photonic band gap, but it cannot exceed 0.5 for technological reasons. Indeed, the walls between holes could collapse for diameters larger than $0.5 p$. Fig.\ref{fig.1} shows the band diagram obtained in the case $D/p = 0.5$. As it can be seen from the figure, the structure yields to a total TM-photonic band gap for frequencies between $0.321p/(2\pi c)$ and $0.349p/(2\pi c)$. The difference from the calculation of S. Massy {\it et al.} \cite{massy03} is probably the difference between the refraction index that are chosen in both cases. Having shown the existance of a photonic band gap in lithium niobate, we have investigated the means of nanostructuring it.

\section{Experimental}

\begin{figure}[!tb]
        \includegraphics[width=0.45\textwidth]{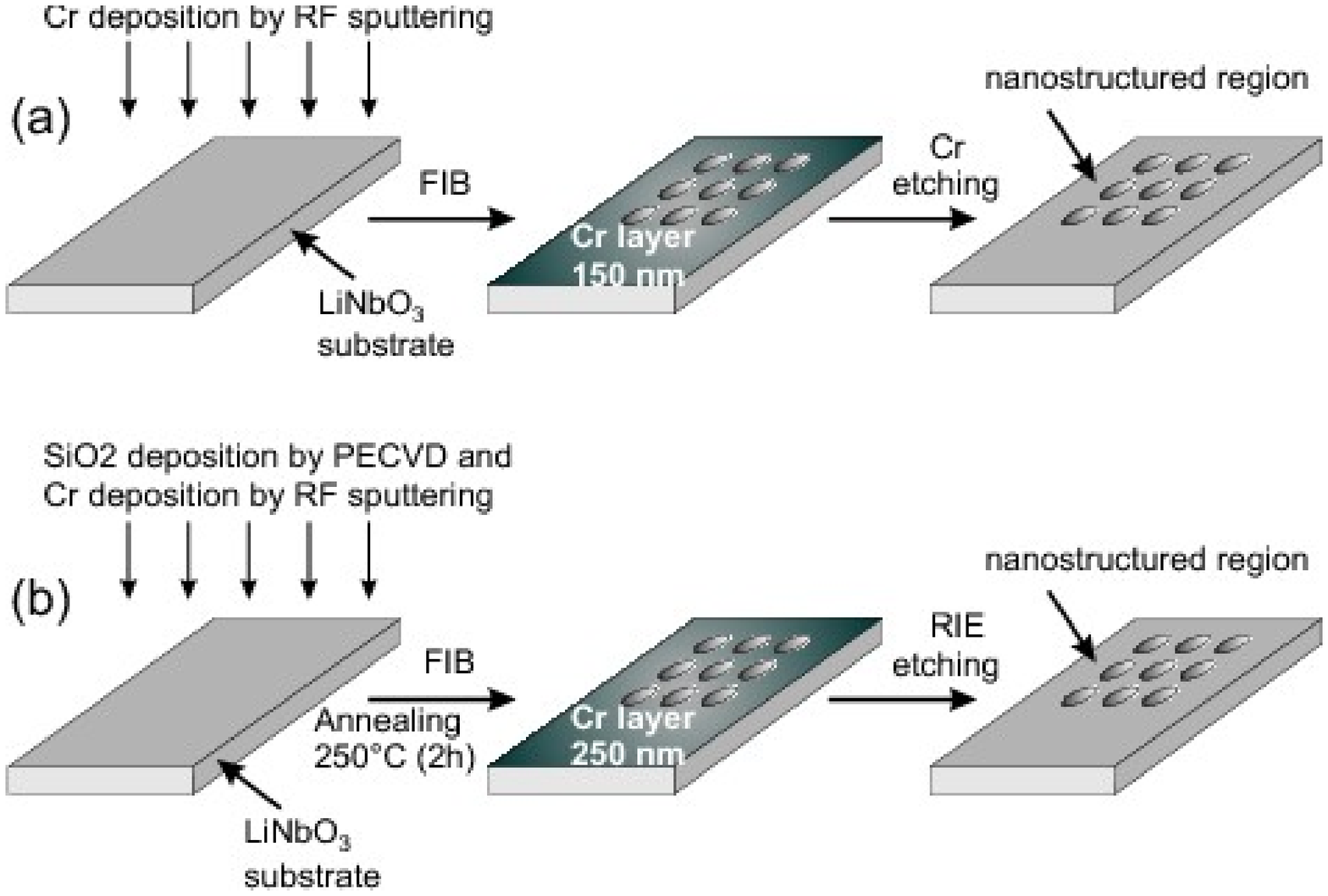}
        \caption{TE and TM band structure for 2D triangular array.}
        \label{fig.2}
\end{figure}

The two fabrication processes are schematically shown in Fig.\ref{fig.2}. The first method -Fig. \ref{fig.2}(a)- is based on a direct etching of the $LiNbO_3$ substrate by FIB milling. The second one -Fig. \ref{fig.2}(b)- uses the FIB to create the metallic mask and the pattern is then transfered to the $LiNbO_3$ substrate by RIE. In both cases the sample area is $1 cm^2$  and the thickness $500 \mu m$. A $Cr$ layer is deposited by electron gun evaporation ({\it Balzers B510}) and grounded with a conductive paste before introduction into the FIB vacuum chamber ($= 2.10^{-6} torr$). In the case of direct FIB writing the thin Cr metal layer ($150nm$) does not modify significantly the etching efficiency. In the second case a thicker $Cr$ layer ($250nm$) is deposited.


The metal-coated substrates are milled using a focused ion beam column {\it Orsay Physics – LEO FIB4400} for the case of FIB milling only -Fig. \ref{fig.2}(a)- and a {\it FEI Dual Beam Strata 235} for the milling of the metallic mask -Fig. \ref{fig.2}(b). This method could be direcly compared with e-beam lithography. The advantage of FIB patterning of the metallic mask is its ability to selectively remove and deposit material without the use of the additional process step of developing a resist layer.

\begin{figure}[!htb]
        \includegraphics[width=0.40\textwidth]{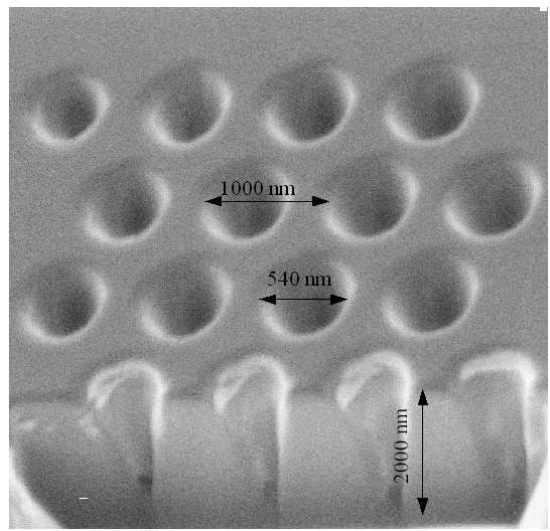}
        \caption{FIB image of the FIB-etched $LiNbO_3$ $4 \times 4$ array of circular
holes.}
        \label{fig.3}
\end{figure}

In the first case (Fig. \ref{fig.2}(a)) we have fabricated an array of $4 \times 4$ circular holes with $540 nm$ diameter  and $1 \mu m$ periodicity. $Ga^+$ ions are emitted with a current of $2 \mu A$ and accelerated by a voltage of $30 kV$. The ions are focused with electrostatic lenses on the sample with a probe current of $66 pA$.  The pseudo-Gaussian-shaped spot size is estimated to be $70 nm$ on the target. The focused ion beam is scanned on the sample by a computer-controlled deflection field to produce the desired pattern ({\it Elphy  Quantum} from {\it Raith}). A FIB-image cross-section of the cavities is shown in Fig. \ref{fig.3}. In order to see the etching depth the sample is tilted  $30^o$ with respect to the FIB axis. As it can be seen from the image, the $4 \times 4$ array exhibits well defined circular holes. The achieved etching depth is approximately $2 \mu m$ and the etching time was 12 minutes. At $1 \mu m$ deep the hole diameter is about $432 nm$. This conical etching shape is due to material redeposition on the sidewalls while milling. In order to reduce the redeposition there are two possible solutions. If the FIB electronics is fast enough ({\it Elphy Quantum} is limited to $300 kHz$) and the spot size small enough one can scan along the hole sidewalls longer and less on the bottom of the pit. A second possibility that we plan to implement in our system is to use gas assisted milling. In particular, $XeF_2$ could help to remove $Nb$ from the etched substrate decreasing then the redeposition.

\begin{figure}[!htb]
        \includegraphics[width=0.40\textwidth]{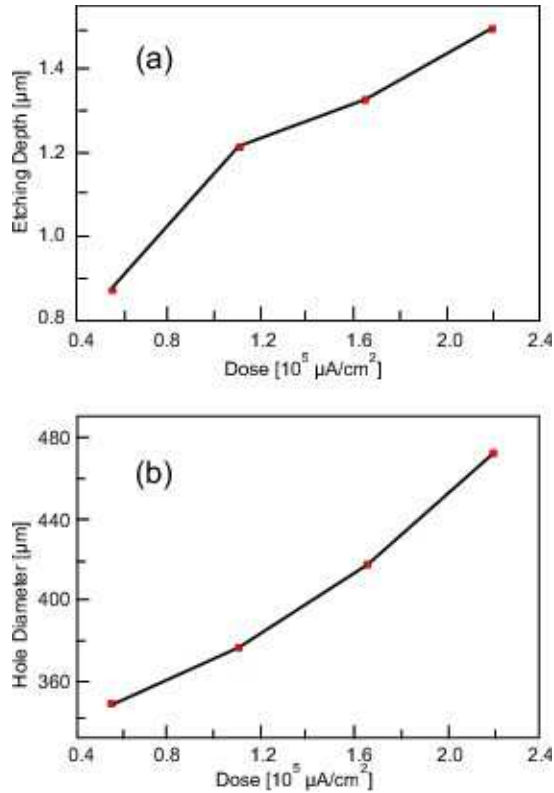}
        \caption{(a) Dependence of the hole depth with respect to the $Ga^{+}$ ion
dose with a cross-section at the first hole-line plane. (b) Dependence of
the hole diameter with respect to the $Ga^{+}$ dose.}
        \label{fig.4}
\end{figure}

\begin{table}[!htb]
\begin{ruledtabular}
\begin{tabular}{cc}
LiNbO3 & Z-cut\\
Pressure & $3 mbar$\\
SF6 flow & $10 sccm$\\
RF power & $150 W$\\
Etching rate & $50 nm/min$\\
$LiNbO_3/Cr$ selectivity & 0.25
\end{tabular}
\end{ruledtabular}
\caption{Parameters of the RIE process.}
\label{Table I}
\end{table}

The etching depth and the real hole diameter can differ from the designed ones and this difference depends on the $Ga^+$ dose. This can be clearly seen in Fig. \ref{fig.4} (a) and (b) where this dependence has been measured. It is clear from the graphs that as the milling time is increased, the etching depth increases but with a subsequent increase in the hole diameter due to beam aberrations. In particular, an increase of almost $30\%$ in the hole diameter and of $40\%$ in the etching depth has been measured as the $Ga^+$ dose is increased by a factor of 4 from the initial value ($4.10^4 \mu A s/cm^2$). Therefore, for practical uses one should start with a designed hole diameter slightly smaller than the desired one.

The second related process requires lower etching time since the desired photonic structure is fabricated at once. In this case, the FIB bombardment is used to pattern a $SiO_2-Cr$ mask previously deposited on the LN substrate, as depicted in Fig. \ref{fig.2}(b). The first step consists in depositing a $100nm$ thick layer of $SiO_2$ by Plasma Enhanced Chemical Vapor Deposition (PECVD). A $250 nm$ thick chrome layer is then deposited on the substrate by sputtering. The  metal is used as a mask for the RIE, while the silica layer prevents the diffusion of chrome into the substrate during the RIE plasma processing and the increase of the optical losses. This layer is not needed in the case of direct FIB milling since the etching is done locally  and the damaged area is defined by the FIB beam size. The samples are annealed at $250^o C$ during 2 hours to release stress. The $SiO_2-Cr$ mask is then nanostructured by FIB patterning, with a current of the sample of $100 pA$. An exposure time of $3.75 s$ is typically required to etch a $250 nm$ diameter circular hole, which is 11 times lower than the one required in the first process.


\begin{figure}[!htb]
        \includegraphics[width=0.40\textwidth]{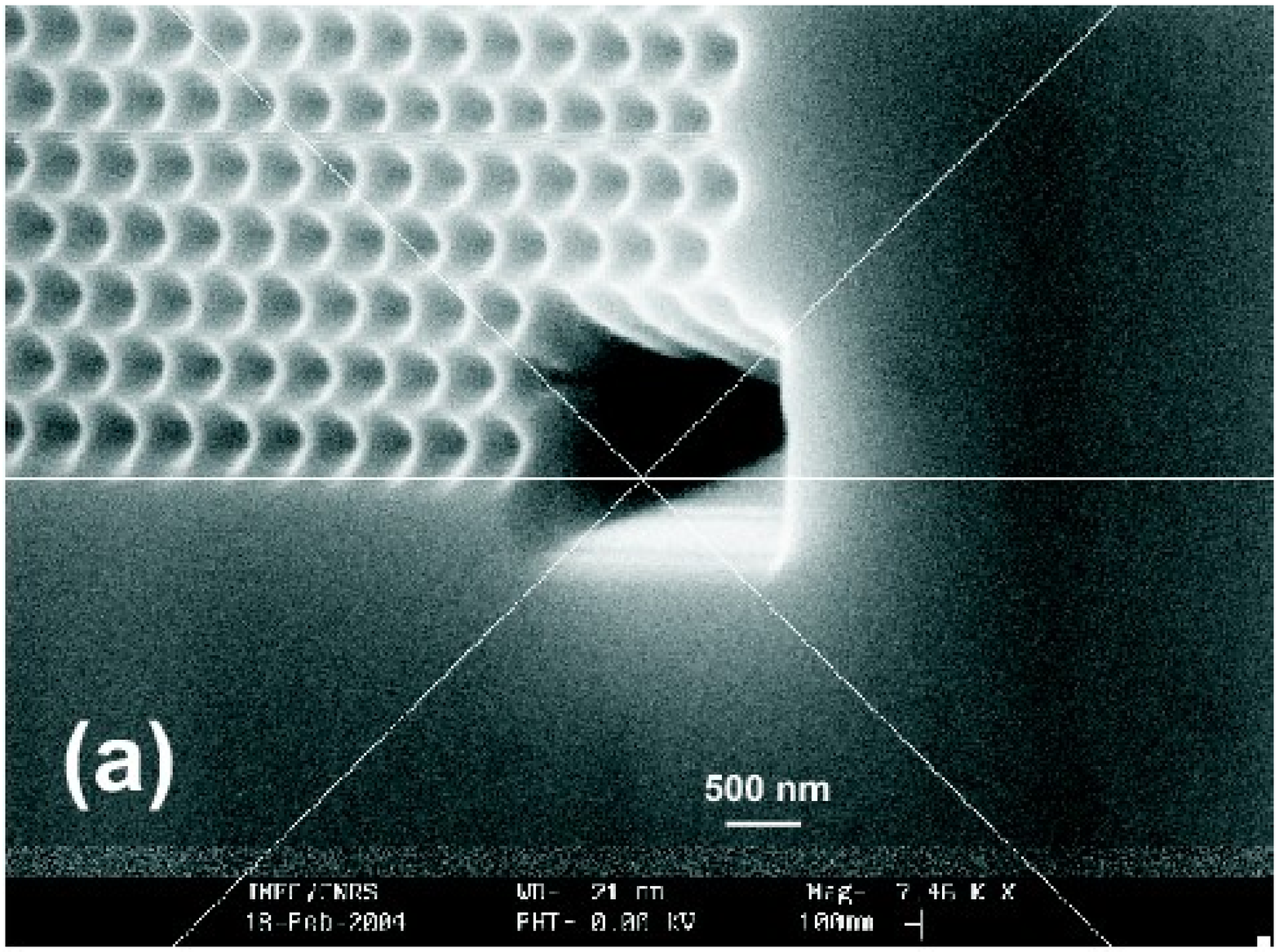}
        \\
        \includegraphics[width=0.40\textwidth]{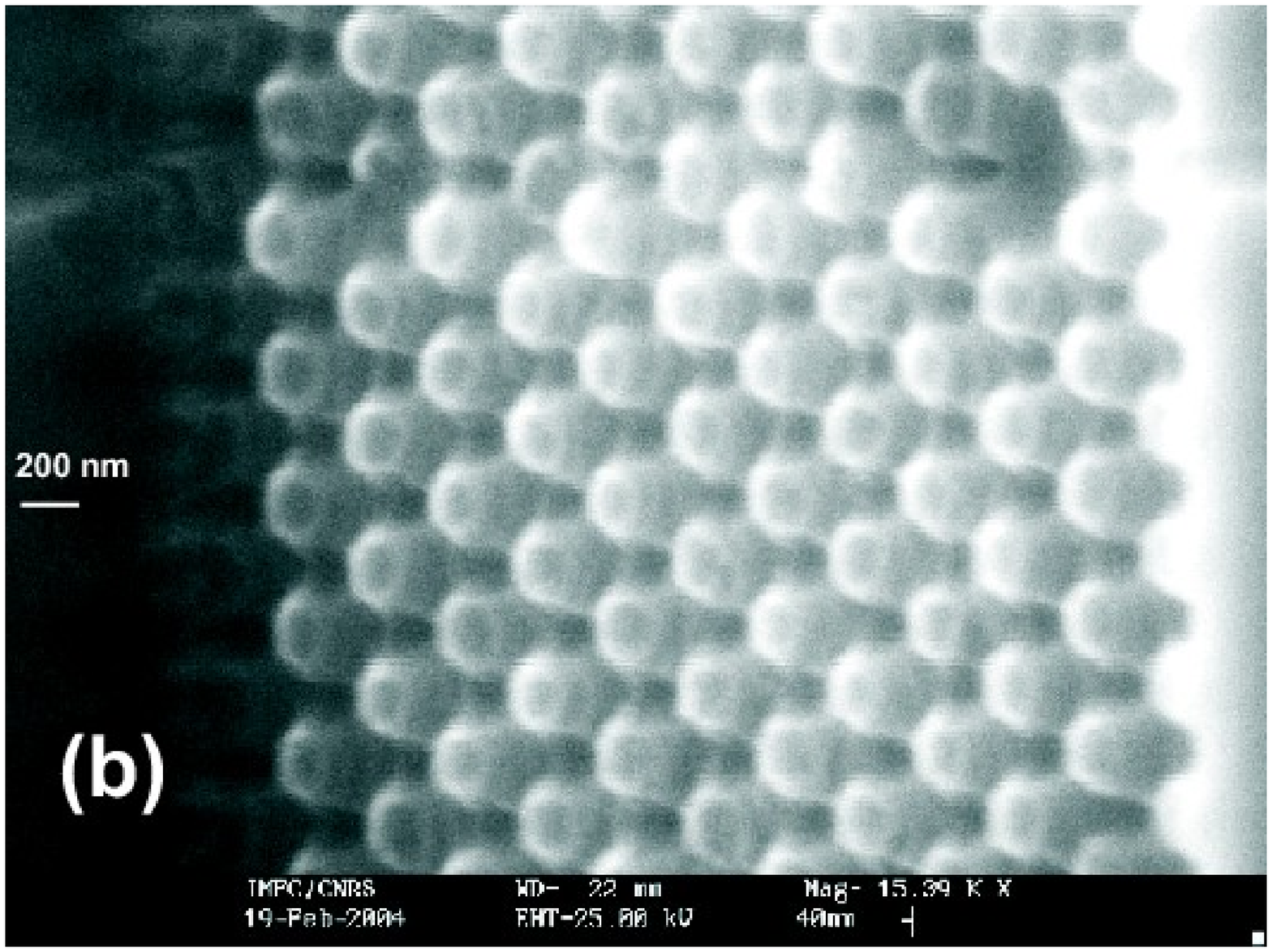}
        \caption{SEM image of the $LiNbO_3$ substrate covered by $250 nm$ of
Chrome after FIB milling and $10min$ of RIE etching. (a) $D = 250 nm$,
(b) $D = 130 nm$.}
        \label{fig.5}
\end{figure}

The pattern (an array of $24 \times 20$ cylindrical holes) is finally transfered to the substrate by RIE. The relevant parameters of this process are detailed in Table \ref{Table I}. It can be noticed that this process requires a very low pressure and a high RF power. In these conditions, the etch rate of the mask is comparable to the etching rate of the substrate. In order to improve the selectivity of etching between the mask and the $LiNbO_3$ substrate, we start the process with an exposition of the target to a $O_2$ ionic plasma (pressure =$100 \mu Bar$ , power = $60 W$). The $250 nm$ thick layer of chrome is then more resistant to the $SF_6$ RIE. The selectivity of the mask is thus estimated to be $1:5$ compared to the $LiNbO_3$ substrate (while the etching selectivity was measured to be of $1:2$ without the $O_2$ ionic plasma). The etching rate of the Z-cut substrate is measured to be $50nm/min$. This process is applied to fabricate a triangular lattice of holes with $D=250nm$ and $D=130nm$ diameters and $p=2 D$ periodicity. Fig. \ref{fig.5} (a) and (b) exhibit the SEM images of the holes after FIB milling and $10min$ of RIE etching. Fig. \ref{fig.5}(a) shows holes with good reproducibility. The etching depth is measured to be $500 nm$. Fig. \ref{fig.5}(b) shows that the $130 nm$ diameter holes were transformed into $130 nm$ diameter rods after RIE etching, while the $250 nm$ diameter holes were well preserved. This is due to a higher etching rate along the sides of the triangular lattice than in the triangle center when the holes are very close to each other. We can infer from these results that the fabrication of small holes ($D \le 200nm$) requires lower RF-power to preserve the initial feature.

\section{Conclusion}
In this work we have presented two alternative techniques to fabricate submicrometric patterns in $LiNbO_3$ by means of a FIB milling or a combination of FIB milling and RIE etching. In particular, $500nm$-diameter circular holes were etched by FIB milling obtaining an etching depth of $2 \mu m$. Material redeposition starts to be a problem for etching depths larger than $1 \mu m$ A $24 \times 20$ array of $250 nm$-diameter circular holes and $500 nm$ periodicity was realized using an alternative method in which the metallic mask is fabricated by FIB milling and the $LiNbO_3$ etching is obtained by $SF_6$ RIE. In this case, the etching depth in $LiNbO_3$ is around $500 nm$ and is limited by the metallic thickness of the mask. Work is in progress to optimize the RIE etching in view of obtaining a higher etching depth. An optical characterization of the patterns is also in progress.

As opposed to the process based on ferro-electric domain inversion, the presented methods are more suitable for implementing nanostructures on both X-cut and Z-cut substrates.

We are currently working on the theoretical analysis and the fabrication of active $LiNbO_3$ 2D photonic crystals. In order to operate at $1.55 \mu m$ our calculations show that we must have a triangular geometry of cylindrical holes of $259 nm$ diameter and $519 nm$ periodicity.
\\
\\
\begin{acknowledgments}
The authors want to thank Jacques Vendeville (FEMTO-ST, Universit\' e de Franche-Comt\' e, Besan\c con) and Elo\"\i se Devaux (Laboratoire des Nanostructures, ISIS, Universit\' e Louis Pasteur, Strasbourg) for technical assistance and Jean-Yves Rauch (FEMTO-ST, Universit\' e de Franche-Comt\' e, Besan\c con) for fruitful discussions.
\end{acknowledgments}

\end{document}